\documentclass[aps,preprint]{revtex4}
\usepackage{amsfonts}
\usepackage{amsmath}
\usepackage{amssymb}
\usepackage{graphicx}

\input{tcilatex}
\begin{document}

\preprint{}
\title[ ]{ Effect of NUT parameter on the analytic extension of the Cauchy
horizon that develop in colliding wave spacetimes}
\author{Ozay Gurtug}
\email{ozay.gurtug@emu.edu.tr}
\affiliation{Department of Physics, Eastern Mediterranean University, G. Magusa, north
Cyprus, via Mersin 10, Turkey.}
\author{Mustafa Halilsoy}
\email{mustafa.halilsoy@emu.edu.tr}
\affiliation{Department of Physics, Eastern Mediterranean University, G. Magusa, north
Cyprus, via Mersin 10, Turkey.}

\begin{abstract}
The Cauchy horizon forming colliding wave solution due to Chandrasekhar and
Xanthopoulos (CX) has been generalized by inclusion of the NUT ( Newman -
Unti - Tamburino) parameter. This is done by transforming the part of the
inner horizon region of a Kerr-Newman-NUT black hole into the space of
colliding waves. By taking appropriate combination of Killing vectors and
analytically extending beyond the Cauchy horizon the time-like hyperbolic
sigularities are resolved as well. This provides another example of its kind
among the type - D metrics with special emphasis on the role of the NUT
parameter. Finally, it is shown that horizons of colliding higher
dimensional plane waves obtained from the black p-branes undergoes a similar
procedure of analytic extension.
\end{abstract}

\maketitle

\section{Introduction}

Black holes (BHs) and gravitational waves (GWs) are the most important
predictions of the Einstein's theory of general relativity. In the history
of these subjects; the discovery of Schwarzchild BH is followed by the
extension to the electrically charged version, Reissner-Nordstr\"{o}m BH.
This is followed by the discovery of rotating versions, Kerr and Kerr-Newman
BHs respectively. On the other side, the topic of GW is developed and its
characteristic properties are studied extensively. One of the research
branch of GWs is to consider their nonlinear interactions which is known in
the literature as colliding gravitational waves (CGWs). This subject was
popular in 1980's and its peculiar features are very well investigated and
catalogued in \cite{JB}. The most common feature is the occurence of
curvature singularities; when two plane GW collide they focus each other in
a way that after a finite time from the instant of the collision, a
spacetime singularity develops. In contrast to this, exceptional solutions
were found which do not exhibit curvature singularities, instead they
develop Killing - Cauchy horizons \cite{CX2}. Under what conditions the
collision of waves produce a space-time curvature singularity or a Killing -
Cauchy horizon ?. To the best of our knowledge this important question
remains still open. The formation of horizons is a common feature for BHs
and CGWs. Although they differ in global structure, they display some
similarities against perturbations, for details see \cite{Y2} and references
contained therein. Yurtsever \cite{Y1} argued that all Killing - Cauchy
horizons that develop in the collision of gravitational waves are unstable
against plane symmetric perturbations. However, it is still early to
generalize the Yurtsever's results whether all horizons formed in CGW are
unstable against arbitrary perturbations.

Plane waves in general relativity are obtained by an infinite boost, i.e. as
the particle speed reaches the speed of light. Collision of plane waves
therefore, carries imprints from the high energy particle collisions; a
popular subject nowadays in connection with the Large Hadron Collision
experiments to be conducted at CERN. Since the Hawking's seminal
contribution, quantum mechanical treatment of space time is believed to
create mini BHs. Apart from the quantal feats, by geometric construction,
classical collision of high energetic particles/wave packets do create also
BHs. This amounts all to the formation of a stable horizon capable to hide
highly concentrated matter inside against instability, which is followed by
an immediate decay. By taking the idea of particle collisions to very
extreme it can be argued that upon mutual focussing strong, finely tuned
non-vacuum plane waves can produce BHs at a classical level as well. This,
however, must be valid for the class of collisions in which a stable Cauchy
horizon forms.

These two topics, BHs and CGWs have developed along their own track.
However, an interesting duality relation between them was observed by
Chandrasekhar and Xanthopoulos (CX) \cite{CX2} ( hereafter this will be
referred as Paper I ). The remarkable aspect of Paper I is that, the two
seemingly unrelated topics BHs and CGWs is shown to be locally isometric.
The solution in Paper I describes the collision of impulsive gravitational
waves accompanied with shock gravitational waves which is locally isometric
to the part of the region in between the inner (Cauchy) and outer (event)
horizons of the Kerr black hole. Later on, the Einstein - Maxwell extension
of Paper I is given by CX\cite{CX1} ( hereafter this will be referred as,
Paper II ). This latter solution is locally isometric to the part of the
trapped region of the Kerr - Newman (KN) black hole. The physical properties
of Paper I and II are analysed in detail by CX. One of the property is the
analytic extension of the space-time beyond the horizon for both Papers I
and II. Such extensions of space-times beyond the horizon are known to be
non-unique and in the extended domain two - dimensional time-like
singularities along hyperbolic arcs are displayed. Another BH related CGW
solution is obtained by Yurtsever \cite{Y3} for parallel polarization ( or
independently by Ferrari - Ibanez\cite{FI}) which is known to be isometric
to the part of interior region of the Schwarzchild BH.

The reserch in the BH area has extended by adding different kind of sources
like cosmological constant, a NUT and an acceleration parameters. BH
solutions with these extensions are of type -D class. The general class of
type - D metrics presented long ago by Plebanski and Demianski (PD) \cite{PD}
that contain all these parameters. Recently, higher dimensional
generalisations of Kerr - NUT -(anti)deSitter spacetime is presented in \cite%
{CLP1, CLP2}. However, the physical interpretation of these parameters
especially the NUT parameter has not been clarified. In general, the NUT
parameter is associated with the gravomagnetic monopole parameter of the
central mass, but the common consensus on its exact physical meaning has not
been attained yet. The physical significance of the NUT parameter in the
Kerr - NUT -(anti)deSitter spacetimes is investigated for the basic four
dimensional case in \cite{GP}. In this article, it is clarified that the
dominance of the NUT parameter over the rotation parameter leaves the
spacetime free of curvature singularities and the corresponding solution is
named as NUT-like solution. But if the rotation parameter dominates the NUT
parameter, the solution is Kerr-like and a ring curvature singularity arise.
Note that, this kind of behaviour on the singularity structure is
independent of the presence of the cosmological constant. In another study 
\cite{BH}, the NUT parameter is interpreted physically as the twist of the
universe.

Our interest in this paper relies completely on the isometric equivalence of
the extended - BH solutions with CGW metrics. For example, cosmological term
in BH solution is equivalent to couple matter fields in the corresponding
isometric CGW problem. In other words, null-shells must be added to the
colliding waves to obtain a cosmological constant. However, the inclusion of
null-shells transform the Cauchy horizon into a curvature singularity.
Therefore, it is of interest to address various subclasses of PD - family of
solutions that admit \ Cauchy horizons in the isometric CGW problems and
investigate further through an analytic continuation beyond the Cauchy
horizon. One such case was identified before by Papacostas and Xanthopoulos
(PX) \cite{PX} who showed that the metric can be extended to full
regularity. Our main purpose in this paper is to identify another case which
incorporates the NUT parameter and to investigate its physical effect in the
space of colliding waves. Our starting point is the Kerr-Newman-NUT (KNN)
black hole with its transform to the CGW. This will provide a NUT extension
of the CX metrics ( both for Paper I and II) which was not identified \
explicitly in Ref. \cite{PX}. We shall verify that our metric is distinct
from the one studied by PX, although both are naturally within the
transforms of PD family.

Although isometry between black holes and CGW hardly finds application
beyond 5-dimensional black holes, black p-branes of arbitrary dimensions
come to our rescue. In this regard we consider colliding higher dimensional
forms obtained from the black p-branes. Recently, we have obtained such
regular solutions \cite{HHU} and in this paper, we show how their analytic
extension works in analogy with the lower dimensions. We show that the
passive role of the higher dimensions does not effect the analytic extension
which is already known to be non-unique.

The paper is organized as follows. In section II, we review the solution and
the analytic extension as explained in paper I. In section III, we present
the NUT extension of paper II and compare it with the PX solution. Section
IV, is devoted for the analytic extension of the metric obtained in section
III. Colliding waves in higher dimensions is considered in section V. The
paper ends with the conclusion in section VII.

\section{A Brief Review of the CX Solution.}

\subsection{The Solution.}

CX found a colliding wave solution in the vacuum Einstein theory which is
locally isometric to the part of the region in between the inner and outer
horizons of the Kerr black hole. The metric is given by

\begin{equation}
ds^{2}=\frac{4Xdudv}{\sqrt{1-u^{2}}\sqrt{1-v^{2}}}-\Delta \delta \frac{X}{Y}%
(dx^{2})^{2}-\frac{Y}{X}(dx^{1}-q_{2}dx^{2})^{2},
\end{equation}%
in which we define coordinates $(\tau ,\sigma )$ (in place of $\eta ,\mu $
coordinates of CX) in terms of the null coordinates $(u,v)$ by

\begin{align}
\tau & =u\sqrt{1-v^{2}}+v\sqrt{1-u^{2}}, \\
&  \notag \\
\sigma & =u\sqrt{1-v^{2}}-v\sqrt{1-u^{2}},  \notag
\end{align}%
and $\Delta =1-\tau ^{2},\delta =1-\sigma ^{2}$. The metric functions are

\begin{align}
X& =(1-p\tau )^{2}+q^{2}\sigma ^{2}, \\
&  \notag \\
Y& =1-p^{2}\tau ^{2}-q^{2}\sigma ^{2}=p^{2}\Delta +q^{2}\delta ,  \notag \\
&  \notag \\
q_{2}& =\frac{2q}{p(1+p)}-\frac{2q\delta (1-p\tau )}{pY},  \notag
\end{align}%
in which the constants $p$ and $q$ must satisfy $p^{2}+q^{2}=1$. We note
that the freedom of overall scaling by a constant is employed throughout the
paper whenever appropriate.The remarkable property of the metric (1) is that
in contrast to other solutions the Killing - Cauchy horizon develops in the
region of interaction.

The metric (1) transforms into the Boyer - Lindquist form of the Kerr black
hole, if the following transformation is used:

\begin{equation}
t=m(x^{1}-\frac{2q}{p(1+p)}x^{2}),\text{ \ \ \ }\phi =\frac{m}{\sqrt{%
m^{2}-a^{2}}}x^{2},\text{ \ \ \ }\tau =\mp \frac{(m-r)}{\sqrt{m^{2}-a^{2}}},%
\text{ \ \ \ }\sigma =\cos \theta ,
\end{equation}%
with

\begin{equation*}
p=\mp \frac{\sqrt{m^{2}-a^{2}}}{m},\text{ \ \ \ }q=\pm \frac{a}{m},\text{\ \
\ \ }(m^{2}>a^{2}).
\end{equation*}%
We have the correspondence,

\begin{equation*}
1-p\tau =\frac{r}{m}\text{ \ \ \ \ and \ \ \ \ }1-\tau ^{2}=-\frac{%
\widetilde{\Delta }}{m^{2}-a^{2}},
\end{equation*}%
where $\widetilde{\Delta }$ now stands for the \textquotedblleft horizon
function\textquotedblright

\begin{equation*}
\widetilde{\Delta }=r^{2}-2mr+a^{2}=(r-r_{-})(r-r_{+}).
\end{equation*}%
With these substitutions the line element (1) is expressed in the form

\begin{equation}
m^{2}ds^{2}=\left( \frac{\widetilde{\Delta }-a^{2}\delta }{\rho ^{2}}\right) %
\left[ dt+\frac{2amr\sin ^{2}\theta }{\widetilde{\Delta }-a^{2}\delta }d\phi %
\right] ^{2}-\frac{\rho ^{2}}{\widetilde{\Delta }}\left[ dr^{2}+\widetilde{%
\Delta }d\theta ^{2}\right] -\left[ \frac{\widetilde{\Delta }\rho ^{2}\sin
^{2}\theta }{\widetilde{\Delta }-a^{2}\delta }\right] d\phi ^{2}
\end{equation}%
where $\rho ^{2}=r^{2}+a^{2}\cos ^{2}\theta ,$ and the constants $a$ and $m$
stand for the parameters of rotation and mass respectively. The roots of $%
\widetilde{\Delta }$, namely, $r_{+}$ and $r_{-\text{ \ }}$are known as the
event (outer) and Cauchy (inner) horizons, respectively. Therefore, the
colliding wave solution is locally isometric to the part of the Kerr metric
in between the two horizons.

\subsection{Analytic Extension Across the Killing - Cauchy Horizon.}

This is done by first observing that the Killing vector $\frac{\partial }{%
\partial x^{2}}$ and its scalar product with the other Killing vector $\frac{%
\partial}{\partial x^{1}}$ vanishes on the Killing - Cauchy horizon at $%
u^{2}+v^{2}=1$ (or $\tau=1$). Therefore, the Killing vector $\frac{\partial 
}{\partial x^{2}}$ becomes null when $u^{2}+v^{2}=1$ (or $\tau=1$).

The non-unique extension of the resulting space-time in the region of
interaction is obtained by the following trasformation.

\begin{equation*}
\xi =se^{\frac{x^{2}}{q}},\text{ \ \ \ \ }\zeta =\emph{se}^{-\frac{x^{2}}{q}}
\end{equation*}%
such that $\xi \zeta =s^{2}=(1-u^{2}-v^{2})^{2}=\Delta \delta $.

With this transformation it was shown by CX that the analytic extension of
the space-time across the Killing - Cauchy horizon reveals the occurence of
time-like curvature singularities along the two dimensional hyperbolic arcs
as shown in Fig.1.

Another interesting colliding wave solution of CX is the one obtained in the
Einstein - Maxwell theory (Paper II). Impulsive gravitational waves and
accompanying gravitational and electromagnetic shock waves are the waves
that takes part in the latter collision and as a result, a non-singular
Killing - Cauchy horizon develops. It was shown also that the interaction
region is locally isometric to the part of the region in between the inner
(Cauchy) and outer (event) horizons of the KN black hole. The same
methodology has also been applied to extend the resulting metric across this
horizon. Their analyses proved that the time-like singularities along
hyperbolic arcs developed in the extended domain in analogy with the
singularity obtained in Paper I.

\section{A New Extension of the CX Colliding Wave Solution.}

In the previous section an isometry has been established between the two
horizons of a Kerr black hole and CGW. The horizon function $\widetilde{%
\Delta }$ admitted two roots: The outer horizon $r_{+}$ and the inner
horizon $r_{-}$. For $r_{-}<r<r_{+}$ we have a case in which the time-like
coordinate becomes space-like ( and vice versa ) and the geometry admits two
space-like Killing vectors instead of one time-like and one space-like. This
particular region can suitably be mapped into the interaction region of CGW.
More generally $\widetilde{\Delta }$ can be a higher order polynomial
admitting more than two roots. No multi-horizon case, however,\ guarantees a
priori any association with CGW. Each one of such cases should be checked
whether this is possible or not. Another particular case where we have still
a quadratic $\widetilde{\Delta },$ which we shall elaborate on, is the KNN
black hole \cite{SK}. The metric describing the KNN \ black hole is given by,

\begin{equation}
ds^{2}=\frac{U^{2}}{\rho ^{2}}\left( dt-Pd\phi \right) ^{2}-\frac{\sin
^{2}\theta }{\rho ^{2}}\left[ \left( F+l^{2}\right) d\phi -adt\right] ^{2}-%
\frac{\rho ^{2}}{U^{2}}dr^{2}-\rho ^{2}d\theta ^{2}
\end{equation}%
where

\begin{align*}
F& =r^{2}+a^{2},\text{ \ \ }U^{2}=r^{2}-2mr+a^{2}+Q^{2}-l^{2},\text{ \ \ }%
\rho ^{2}=r^{2}+\lambda ^{2}, \\
& \\
P& =a\sin ^{2}\theta -2l\cos \theta ,\text{ \ \ \ }\lambda =l+a\cos \theta ,
\end{align*}%
in which the constants $a,Q$ and $l$ stand for the rotation, electric charge
and NUT parameters, respectively. It describes a stationary axially
symmetric body with gravitomagnetic monopole and dipole moments associated
with nonzero values of the NUT $\left( l\right) $ and Kerr $\left( a\right) $
parameters. It can also be interpreted as the gravitational dyon solution
that represents the gravitational field of a rotating body having both
gravitational electric and magnetic charges. The KNN solution belongs to the
Petrov type-D solutions of the Einstein-Maxwell equations for which the
space-time admits separable Hamilton-Jacobi, Klein-Gordon and Dirac
equations, \cite{DT},\cite{MD}. One remarkable property satisfied by the
Kerr - NUT metric ( i.e. for $Q=0$ ) is the duality invariance which is
defined by $r\leftrightarrow i\lambda $ and $m\leftrightarrow il$ \cite{MD}.
Is there a similar duality invariance associated with the CGW metric ?. The
mixing of radial and angular coordinates has already been imposed by the
isometry transformation ( see Eq.(7) below) and any further symmetry is
prohibited by the symmetry of the CGW. As a matter of fact the
transformation is itself a duality, i.e. the \textquotedblleft CX -
duality\textquotedblright\ between black holes and CGWs. Once we are in the
space of CGW the invariance is restricted to $u\leftrightarrow v$ in the
null plane and to $x\leftrightarrow y$ in the orthogonal space. Any further
mixing of these two sectors does not work.

Having in mind all these features, it would be interesting \ to find the
corresponding isometric geometry in the space of CGW and explore its
physical properties. The corresponding CGW metric can be determined by using
the transformation

\begin{equation}
r=m+\sqrt{m^{2}+l^{2}-a^{2}-Q^{2}}\tau ,\text{ \ \ }\sigma =\cos \theta ,%
\text{ \ \ }t=x,\text{ \ \ }\phi =y,
\end{equation}%
and after setting the mass parameter $m=1$, the CGW metric can be cast in
the form

\begin{equation}
ds^{2}=X\left( \frac{d\tau ^{2}}{\Delta }-\frac{d\sigma ^{2}}{\delta }%
\right) -X^{-1}\left( Rdx^{2}+Edy^{2}-2Gdxdy\right) .
\end{equation}%
Our notation and abbreviations in this metric are as follow

\begin{eqnarray}
X &=&\left( p+\tau \right) ^{2}+\left( l_{0}+a_{0}\sigma \right)
^{2}=B-a_{0}A, \\
&&  \notag \\
R &=&\Delta +a_{0}^{2}\delta ,\text{ \ \ \ \ \ \ \ \ \ }E=\Delta
A^{2}+\delta B^{2},\text{ \ \ \ \ }G=\Delta A+a_{0}\delta B,  \notag \\
&&  \notag \\
\text{ \ \ \ \ }A &=&a_{0}\delta -2l_{0}\sigma ,\text{ \ \ \ \ }B=\left(
p+\tau \right) ^{2}+a_{0}^{2}+l_{0}^{2},  \notag \\
&&  \notag
\end{eqnarray}%
where

\begin{eqnarray}
\Delta &=&1-\tau ^{2},\text{ \ \ \ \ }\delta =1-\sigma ^{2}, \\
&&  \notag \\
\tau &=&\sin \left( \widetilde{a}u+\widetilde{b}v\right) ,\text{ \ \ \ \ }%
\sigma =\sin \left( \widetilde{a}u-\widetilde{b}v\right) ,\text{ \ \ \ \ and
\ \ \ \ \ }\widetilde{a},\widetilde{b}=\text{ \textit{constant},}  \notag \\
&&  \notag \\
l_{0} &=&pl,\text{ \ \ \ }a_{0}=ap,\text{ \ \ \ }q=Qp,  \notag \\
&&  \notag
\end{eqnarray}%
in which $p=\frac{1}{\sqrt{1+l^{2}-a^{2}-Q^{2}}}$\ so that $%
p^{2}+l_{0}^{2}-a_{0}^{2}-q^{2}=1.$ Obviously it is clear from the square
root expression that the NUT parameter $l$ is constrained by $%
1+l^{2}>a^{2}+Q^{2}$. This constraint condition is crucial for having a
non-singular CGW space-time. In order to interpret the foregoing metric as
CGW the null coordinates must be multiplied everywhere by the unit step
functions, i.e. $u\rightarrow u\theta \left( u\right) $ and $v\rightarrow
v\theta \left( v\right) $. For the sake of simplicity in the rest of the
paper we shall set $\widetilde{a}=\widetilde{b}=1$. Let us note also that
the $\left( \eta ,\mu \right) $ coordinates of CX are equivalent to our $%
\left( \tau ,\sigma \right) $ by the additional transformations $%
u\rightarrow \sin u$ and $v\rightarrow \sin v$. Although this rescaling of
coordinates is not imperative it has the advantage that the $g_{uv}$
component of the metric can be expressed in the simplest possible form,
namely $g_{uv}=X$ ( after an overall scaling by a factor 2 ). Note that for $%
Q=0,$ the metric in Eq.(8) overlaps with the one obtained long ago by Wang 
\cite{AW} which is the NUT extension of Paper I. However, the physical
significance of the NUT parameter was not the scope of that paper.

The non zero Ricci components due to the em field in the Newman - Penrose
formalism are given, after tedious calculation by%
\begin{eqnarray}
\Phi _{22} &=&\theta (u)q^{2}X^{-1}, \\
&&  \notag \\
\Phi _{00} &=&\theta (v)q^{2}X^{-1},  \notag \\
&&  \notag \\
\Phi _{02} &=&\theta (u)\theta (v)q^{2}X^{-1}\left( ER\right) ^{-1/2}\left[ 
\sqrt{\Delta \delta }\left( B+a_{0}A\right) +i\left( \Delta A-a_{0}\delta
B\right) \right] .  \notag
\end{eqnarray}%
We recall that exact determination of the phases $(f,g)$ of the Maxwell
spinors $\Phi _{2}=qX^{-1/2}e^{if}$ and $\Phi _{0}=qX^{-1/2}e^{ig}$ is
possible through the tedious integration of the Maxwell equations.
Fourtunately this is not necessary in the present study and for this reason
we shall ignore it.

\subsection{Initial Data and The Boundary Conditions.}

In order that the metric (8) can be interpreted as a CGW metric it has to
satisfy certain boundary conditions. To determine the initial data we
extrapolate the interaction region back a$^{\prime }$la CX to find the
incoming regions. The initial data associated with Region II $\ (u\geq
0,v<0) $ is obtained by dropping the $v$ in the metric. This amounts to the
substitution $\tau =\sigma =\sin (u\theta \left( u\right) )$, so that the
metric functions take the form

\begin{eqnarray}
R(u) &=&\left( 1+a_{0}^{2}\right) \cos ^{2}u, \\
&&  \notag \\
X(u) &=&\left( p+\sin u\right) ^{2}+\left( l_{0}+a_{0}\sin u\right) ^{2}, 
\notag \\
&&  \notag \\
E(u) &=&\cos ^{2}u\left[ \left( a_{0}\cos ^{2}u-2l_{0}\sin u\right)
^{2}+\left( \left( p+\sin u\right) ^{2}+a_{0}^{2}+l_{0}^{2}\right) ^{2}%
\right] ,  \notag \\
&&  \notag \\
G(u) &=&\cos ^{2}u\left[ a_{0}\cos ^{2}u-2l_{0}\sin u+a_{0}\left( \left(
p+\sin u\right) ^{2}+a_{0}^{2}+l_{0}^{2}\right) \right] ,  \notag
\end{eqnarray}%
in which$\ u$ is implied with the unit step function. The only non zero
Ricci component in this region is given by

\begin{equation}
\Phi _{22}(u)=\theta (u)q^{2}X^{-1}(u),
\end{equation}%
while the gravitational wave component $\Psi _{4}\left( u\right) $ has the
form

\begin{equation}
\Psi _{4}\left( u\right) =\left( \text{\textit{constant}}\right) \delta
\left( u\right) +\theta \left( u\right) L\left( u\right) ,
\end{equation}%
in which $\delta \left( u\right) $ stands for the Dirac delta function and $%
L\left( u\right) $ is a well defined function. An impulsive gravitational
wave superposed with a shock gravitational wave constitutes the general
feature of these waves. Similar data is obtained for region III $\left(
u<0,v\geq 0\right) $ by the substitution $\tau =-\sigma =\sin (v\theta
\left( v\right) )$, which will not be given. Obviously region III has the
corresponding non-vanishing Ricci and Weyl components

\begin{eqnarray}
\Phi _{00}(v) &=&\theta (v)q^{2}X^{-1}(v), \\
&&  \notag \\
\Psi _{0}\left( v\right) &=&\left( \text{\textit{constant}}\right) \delta
\left( v\right) +\theta \left( v\right) K\left( v\right) ,  \notag
\end{eqnarray}

where the function $K(v)$ similar to $L\left( u\right) $ above is a well
defined function whose exact form is not of much interest here and we shall
not give them explicitly. Expectedly all these components are much more
involved relative to the ones considered by CX.

Further extrapolation, by letting $u<0$ and $v<0$ in (8) reduces our metric
into

\begin{equation}
ds^{2}=4\left( p^{2}+l_{0}^{2}\right) dudv-\left( dx-a_{0}dy\right)
^{2}-\left( cdy-a_{0}dx\right) ^{2}\text{ ,\ \ \ \ \ }
\end{equation}%
where

\begin{equation*}
c=p^{2}+a_{0}^{2}+l_{0}^{2},
\end{equation*}%
which is manifestly flat in a rescaled coordinate system. It is evident from
the data (Eq. 13 and 14 ) for region II ( and similar expression for region
III ) that both of the parameters $a_{0}$ and $l_{0}$ corresponding to the
angular momentum and NUT parameters of the original black hole solution
transform under the isometry (7) into the parameter of cross ( or second )
polarization of the waves. Linear polarization limit corresponds to the case 
$l_{0}=0=a_{0}$, as can easily be checked from the metric component $g_{xy}$.

In the presence of electromagnetic waves it had been shown that the
appropriate boundary conditions are those of O'Brien and Synge \cite{OBS}.
We adopt here the same boundary conditions, provided we verify the absence
of any current sources on the null boundaries. These can be summarized by
the continuity requirements of $g_{\mu \nu }$, $g^{ij}g_{ij,0\text{ }}$,
where $\left( i,j=1,2,3\right) $ and $x^{0}$ refers to the null coordinates
with the condition that $g_{00}=0$. For more detail on the choice of the
boundary conditions we refer to Ref. [1] and references cited therein. While
the other conditions are self evident the critical condition to be checked
is the continuity of

\begin{equation*}
g^{ij}g_{ij,u},\text{ \ \ \ \ \ \ \ \ \ \ \ \ \ \ \ \ on \ \ \ \ \ \ \ \ \ \
\ \ \ \ \ \ \ }u=0
\end{equation*}%
and

\begin{equation*}
g^{ij}g_{ij,v},\text{ \ \ \ \ \ \ \ \ \ \ \ \ \ \ \ \ on \ \ \ \ \ \ \ \ \ \
\ \ \ \ \ \ \ }v=0.
\end{equation*}%
When worked out in detail these reduce to the requirements that $\left( \ln
\Delta \delta \right) _{,u}$ and $\left( \ln \Delta \delta \right) _{,v}$
are both continuous across $u=0$ and $v=0$, respectively. In summary, these
are both continuous, the O'Brien - Synge conditions are satisfied and no
extra sources are created in the collision process derived from the isometry
with the KNN black hole.

\subsection{The Weyl and Ricci Scalars.}

Our interaction region $\left( u>0,v>0\right) ,$ metric (8) is equivalent
(isometric) to the KNN metric (6), in fact it is obtained from the latter by
a coordinate transformation. In this section we wish to make use of this
advantage to find a proper tetrad that gives $\Psi _{2}$ and $\Phi _{11}$ as
the only nonvanishing Weyl and Ricci \ scalars.

Our choice of the proper tetrad is,

\begin{align}
l^{\mu }& =\left( k,0,-\frac{D}{\Delta },-\frac{a}{\Delta }\right) , \\
&  \notag \\
2n^{\mu }& =\frac{1}{k^{2}Z}\left( k\Delta ,0,D,a\right) ,  \notag \\
&  \notag \\
\sqrt{2}m^{\mu }& =\frac{1}{\sqrt{\delta }\left( l+a\sigma -i(1+k\tau
)\right) }\left( 0,i\delta ,a\delta -2l\sigma ,1\right) ,  \notag
\end{align}%
where

\begin{align*}
Z& =\left( 1+k\tau \right) ^{2}+\left( l+a\sigma \right) ^{2}, \\
& \\
D& =\left( 1+k\tau \right) ^{2}+l^{2}+a^{2}, \\
& \\
k^{2}& =1-a^{2}+l^{2}-Q^{2},
\end{align*}%
in which the parameter $k$ is related to our previous parameter $p$ through

\begin{equation*}
p=\frac{1}{k}.
\end{equation*}

In this proper tetrad the type-D character of our space-time becomes
manifest with the $\Psi _{2}$ and $\Phi _{11}$ as follows

\begin{equation}
\Psi _{2}=-\frac{1-il}{\left[ 1+k\tau -i\left( l+a\sigma \right) \right] ^{3}%
},
\end{equation}

\begin{equation}
\Phi _{11}=\frac{Q^{2}}{2\left[ \left( l+a\sigma \right) ^{2}+\left( 1+k\tau
\right) ^{2}\right] ^{2}}.
\end{equation}%
Here although $Q$ is the electric charge of the original black hole, under
the isometric transformation (7) it transforms together with the other
parameters of the black hole, into $q$ of the CGW metric. It is clear to
observe that when $Q=0$, the solution reduces to Kerr - NUT, which is the
NUT extension of the paper I. When $Q=l=0$, our solution reduces to the
vacuum ( i.e. Kerr ) Einstein solution which corresponds to the paper I. If
we further choose $a=0$ but $Q\neq 0$ and $l\neq 0$, the resulting metric
corresponds to the charged - NUT metric which has not been considered by CX.
In general, the term $l+a\sigma $ in the Weyl scalar $\Psi _{2}$ represent
the twist parameter of the gravitational wave. Hence, the NUT parameter $l$
has the tendency to increase the existing twist in the Paper I and II.

Investigation of $\Psi _{2}$ and $\Phi _{11\text{ }}$reveals also that the
NUT extension of paper II is another new non-singular solution. In the limit 
$\tau \rightarrow 1$, both remain finite, implying that the hypersurface $%
\tau =1$ (or $\sin ^{2}u+\sin ^{2}v=1)$ is a Cauchy-horizon instead of a
singular hypersurface.

\subsection{Comparison with the PX metric.}

The PX metric obtained from a member of PD family is expressed, after minor
rearrangements by 
\begin{equation}
ds^{2}=\left( t^{2}+z^{2}\right) \left( \frac{d\tau ^{2}}{\Delta }-\frac{%
d\sigma ^{2}}{\delta }\right) -\frac{1}{\left( t^{2}+z^{2}\right) }\left[
\beta ^{2}\Delta \left( dy-z^{2}dx\right) ^{2}+\alpha ^{2}\delta \left(
dy+t^{2}dx\right) ^{2}\right] .
\end{equation}

The crucial parameters $\alpha >0$ and $\beta >0$ are constants while $%
t=t(\tau )$ and $z=z(\sigma )$ are both linear functions of their arguments
inherited from the PD \ family of solutions without the cosmological
constant. We wish to compare this metric with our metric (8) which was
obtained directly from the KNN black hole. In order to establish a
connection ( if any) between (8) and (20) it is suggestive to identify

\begin{eqnarray}
t &=&p+\tau , \\
z &=&l_{0}+a_{0}\sigma ,  \notag
\end{eqnarray}%
and compare the $(x,y)$ components of both metrics. By choosing $\beta =1$
and $\alpha =a_{0}>0,$ $R$ of (8) equals $g_{yy}$ of (20). Next, by
rescaling $x$ in accordance with $x=-\frac{1}{a_{0}}\overline{x}$ (and
removing the bar over $x$ afterwards) we expect that $(E,G)$ functions will
also match with $g_{xx}$ and $g_{xy}$ of (20). It turns out that such an
identification fails except at the singular hypersurfaces characterized by $%
\tau =1$ and $\sigma =\pm 1.$ This verifies that the metrics (8) and (20)
are different limiting cases of the original PD familiy members. In the
proper tetrad and in terms of the functions $(t,z)$ our Weyl and Ricci
components ( i.e. (18) and (19) ) take the compact forms

\begin{eqnarray}
\Psi _{2} &=&-\frac{p^{2}\left( p-il_{0}\right) }{\left( t-iz\right) ^{3}},
\\
\Phi _{11} &=&\frac{p^{4}Q^{2}}{2\left( t^{2}+z^{2}\right) ^{2}}.  \notag
\end{eqnarray}%
For the metric (20), on the other hand, the choice of the proper null tetrad
one-forms

\begin{eqnarray}
\sqrt{2}l &=&\frac{A}{\sqrt{\Delta }}d\tau -\frac{\sqrt{\Delta }}{A}\left(
dy-z^{2}dx\right) , \\
\sqrt{2}n &=&\frac{A}{\sqrt{\Delta }}d\tau +\frac{\sqrt{\Delta }}{A}\left(
dy-z^{2}dx\right) ,  \notag \\
\sqrt{2}m &=&-\frac{A}{\sqrt{\delta }}d\sigma +ia_{0}\frac{\sqrt{\delta }}{A}%
\left( dy+t^{2}dx\right) ,  \notag \\
A^{2} &=&t^{2}+z^{2},  \notag
\end{eqnarray}%
yields

\begin{eqnarray}
\Psi _{2} &=&\frac{p^{2}Q^{2}}{\left( t^{2}+z^{2}\right) \left( t-iz\right)
^{2}}-\frac{p-il_{0}}{\left( t-iz\right) ^{3}}, \\
\Phi _{11} &=&\frac{p^{2}Q^{2}}{2\left( t^{2}+z^{2}\right) ^{2}}.  \notag
\end{eqnarray}

It is observed that the non-vanishing tetrad scalars of the two metrics are
not identical. This originates from the fact that in the original PD metrics
further identification and linear transformation should be applied before
arriving at the KNN metric. Since this has not been done the NUT parameter
remained unidentified and (8) remained distinct from (20). We observe after
setting $\beta =1$ and $\alpha =a_{0}$ that, the PX metric does not possess
the limit $a_{0}=0$, whereas (8) remains meaningful. The $\Psi _{2}$ in (8)
reveals that all limiting cases are well defined; $l_{0}=0$, $a_{0}\neq 0;$ $%
l_{0}\neq 0$ , $a_{0}=0$; or $l_{0}=0=a_{0}.$ Obviously the charged - NUT
case corresponds to $a_{0}=0$ , $l_{0}\neq 0$ and $Q\neq 0$, and is a new
case not included in the analysis of CX ( or PX). In the special case $Q=0,$ 
$p=1$ (or $l=a$), however the two metrics become identical. One more
distinction arises when we consider an unbounded NUT parameter $%
(l\rightarrow \infty )$. Evidently the metric (8) has $\Psi _{2}=0=\Phi
_{11} $, while for the metric (20) we obtain $\Psi _{2}\neq 0=\Phi _{11}$.

\section{Analytic Extension of the Space-Time Across the Horizon.}

The line element (8) describes the NUT extension of the colliding wave
solution due to the CX in the Einstein - Maxwell theory. The determinant of
the metric in the $u,v,x,y$ coordinates is given by,

\begin{equation}
\mid g\mid =X^{2}\Delta \delta .
\end{equation}

It is obvious that as $\tau\rightarrow1$, the determinant vanishes on the
surface $\tau=1$ which is equivalent to $u+v=\pi/2$. We have seen in the
previous section that the Weyl and Ricci scalars remain finite on this
surface, indicating no scalar curvature singularity. Therefore the vanishing
of determinant would only has meant that there exists a coordinate
singularity and it can be removed by an appropriate transformation.

In order to perform the analytic extension, we shall adopt the method given
by CX in Paper I. As a requirement of the method, at least one of the
Killing vector fields should become null on the horizon. This is provided by
calculating the norm of the Killing vectors on the horizon surface. The norm
of the Killing vectors are

\begin{align}
& \mid \partial _{x}\mid ^{2}=g_{xx}=-X^{-1}R, \\
&  \notag \\
& \mid \partial _{y}\mid ^{2}=g_{yy}=-X^{-1}E.  \notag
\end{align}

We observe that none of the norms of the Killing vectors vanish on the
horizon surface ( as $\tau \rightarrow 1$ ). In other words the Killing
vectors $\partial _{x}$ and \ $\partial _{y}$ are both space-like. As it was
stated in Paper I, the Killing vectors $\partial _{x}$ or \ $\partial _{y}$
can be made null if the integration constant for the metric function $q_{2}$
is chosen properly. In order to assign the same role on these Killing
vectors we take their linear combination. As we shall explore in the
following section this choice plays a crucial role as far as the singularity
structure is concerned in the extended domain.

We choose the new Killing vector as,

\begin{equation}
\xi _{1}^{\mu }=\alpha \xi _{x}^{\mu }+\beta \xi _{y}^{\mu },
\end{equation}%
where $\alpha $ and $\beta $ are non-zero constants. We impose the condition
that the norm of the new Killing vector $\mid \xi _{1}^{\mu }\mid ^{2}$%
should vanish on the horizon surface as $\tau \rightarrow 1$.

\begin{equation*}
\mid \xi _{1}^{\mu }\mid ^{2}=g_{\mu \nu }\left( \alpha \xi _{x}^{\mu
}+\beta \xi _{y}^{\mu }\right) \left( \alpha \xi _{x}^{\nu }+\beta \xi
_{y}^{\nu }\right) =0.
\end{equation*}%
Hence, the new Killing vector is obtained as,

\begin{equation}
\xi _{1}^{\mu }=\delta _{x}^{\mu }+c_{1}\delta _{y}^{\mu },
\end{equation}%
where $c_{1}=\frac{\beta }{\alpha }=\frac{a_{0}}{X_{0}}$ with $%
X_{0}=(1+p)^{2}+a_{0}^{2}+l_{0}^{2}$. Note that the major difference in our
choice is that, the new Killing vector lies in the $xy$ - plane ( $x^{1}=x$
and $x^{2}=y$), whereas, in CX case the Killing vector was becoming null on
the $x^{2}$ ( $=y$\ ) axis.

At this stage we define two new coordinates as

\begin{align}
\overline{x}& =x+c_{1}y, \\
&  \notag \\
\overline{y}& =y-c_{1}x.  \notag
\end{align}%
In terms of the null coordinates, the metric (8) becomes

\begin{align}
ds^{2}& =2Xdudv-\frac{X^{-1}X_{0}^{2}}{\left( X_{0}^{2}+a_{0}^{2}\right) ^{2}%
}\{\left[ X_{0}^{2}R+a_{0}^{2}E-2GX_{0}a_{0}\right] d\overline{x}^{2} \\
&  \notag \\
& +\left[ a_{0}^{2}R+X_{0}^{2}E+2GX_{0}a_{0}\right] d\overline{y}^{2}  \notag
\\
&  \notag \\
& -2\left[ X_{0}a_{0}\left( R-E\right) +\left( X_{0}^{2}-a_{0}^{2}\right) G%
\right] d\overline{x}d\overline{y}\}  \notag
\end{align}%
The norm of the new Killing vector $\partial _{\overline{x}}$ and its scalar
product with the other Killing vector $\partial _{\overline{y}}$ are given
by,

\begin{equation}
\mid \partial _{\overline{x}}\mid ^{2}=-\frac{X^{-1}X_{0}^{2}}{%
X_{0}^{2}+a_{0}^{2}}\left\{ \left[ X+\left( 1-\tau \right) \left( 2p+1+\tau
\right) \right] ^{2}+\frac{a_{0}^{2}\delta \left( 1-\tau \right) }{\left(
1+\tau \right) }\left( 2p+1+\tau \right) ^{2}\right\} \Delta
\end{equation}%
and

\begin{align}
\left( \partial _{\overline{x}}\cdot \partial _{\overline{y}}\right) & =2%
\frac{X^{-1}X_{0}^{2}}{X_{0}^{2}+a_{0}^{2}}\{X_{0}a_{0}+A\left\{ X_{0}\left[
X+\left( 1-\tau \right) \left( 2p+1+\tau \right) \right] -a_{0}^{2}\right\}
\\
&  \notag \\
& +\frac{a_{0}\delta \left( 2p+1+\tau \right) }{1+\tau }\left\{
X_{0}B+a_{0}^{2}\right\} \}\Delta .  \notag
\end{align}%
It is clear to observe that both vanish in the limit $\tau \rightarrow 1$.
We rewrite the metric (30) in terms of the new variables $\widetilde{l}$ and 
$r$ defined by

\begin{align}
\widetilde{l}& =\sqrt{\Delta \delta }=1-\sin ^{2}u-\sin ^{2}v, \\
&  \notag \\
r& =\tau \sigma =\cos ^{2}v-\cos ^{2}u.  \notag
\end{align}%
This transforms metric (30) into

\begin{align}
ds^{2}& =\frac{X}{H}\left( d\widetilde{l}^{2}-dr^{2}\right) -\frac{%
X^{-1}X_{0}^{2}}{(X_{0}^{2}+a_{0}^{2})^{2}}\{\left[
X_{0}^{2}R+a_{0}^{2}E-2GX_{0}a_{0}\right] d\overline{x}^{2} \\
&  \notag \\
& +\left[ a_{0}^{2}R+X_{0}^{2}E+2GX_{0}a_{0}\right] d\overline{y}^{2}  \notag
\\
&  \notag \\
& -2\left[ X_{0}a_{0}\left( R-E\right) +\left( X_{0}^{2}-a_{0}^{2}\right) G%
\right] d\overline{x}d\overline{y}\},  \notag
\end{align}%
where $H=\delta -\Delta =\sin 2u\sin 2v$. It can be shown easily that the
coordinate singularity at $\tau =1$ is removed when we apply the following
transformation,

\begin{equation}
\xi =\widetilde{l}e^{c\overline{x}}\text{ \ \ \ \ and\ \ \ \ \ \ \ \ \ }%
\zeta =\widetilde{l}e^{-c\overline{x}},
\end{equation}%
in which $c=\frac{X_{0}}{X_{0}^{2}+a_{0}^{2}}$\ and in terms of the new
coordinates ( $\xi ,\zeta $ ), we have

\begin{align}
\tau & =\frac{1}{2}\left\{ \sqrt{\left( 1+r\right) ^{2}-\xi \zeta }+\sqrt{%
\left( 1-r\right) ^{2}-\xi \zeta }\right\} , \\
&  \notag \\
\sigma & =\frac{1}{2}\left\{ \sqrt{\left( 1+r\right) ^{2}-\xi \zeta }-\sqrt{%
\left( 1-r\right) ^{2}-\xi \zeta }\right\} ,  \notag \\
&  \notag \\
H& =\sqrt{\left( 1+\xi \zeta -r^{2}\right) ^{2}-4\xi \zeta }.  \notag
\end{align}

The exact metric (30) can be written in such a way that the absence of any
singularity when $\xi =0$ and/or $\zeta =0$ is manifest, as follows

\begin{equation}
ds^{2}=\frac{1}{2HX\delta }\left\{ \overline{A}\left( \zeta ^{2}d\xi
^{2}+\xi ^{2}d\zeta ^{2}\right) +\overline{B}d\xi d\zeta \right\} +C\left(
\zeta d\xi -\xi d\zeta \right) d\overline{y}-\frac{X}{H}dr^{2}-Dd\overline{y}%
^{2},
\end{equation}

where

\begin{align*}
\overline{A}& =\frac{1}{2\delta }\left( X^{2}-\frac{\Sigma }{\left( 1+\tau
\right) ^{2}}\right) , \\
& \\
\overline{B}& =X^{2}\left( 2\delta -\Delta \right) +\frac{1-\tau }{1+\tau }%
\Sigma , \\
& \\
C& =\frac{X_{0}\widetilde{\Sigma }}{X\left( X_{0}^{2}+a_{0}^{2}\right)
\delta }, \\
& \\
D& =\frac{X_{0}^{2}\widetilde{\widetilde{\Sigma }}}{X\left(
X_{0}^{2}+a_{0}^{2}\right) ^{2}},
\end{align*}

with

\begin{align*}
\Sigma & =H\left( 2p+1+\tau \right) \left[ 2X(1+\tau )+\left( 2p+1+\tau
\right) R\right] , \\
& \\
\widetilde{\Sigma }& =a_{0}X_{0}+A\left\{ X_{0}\left[ X+\left( 1-\tau
\right) \left( 2p+1+\tau \right) \right] -a_{0}^{2}\right\} + \\
& \\
& \frac{a_{0}\delta \left( 2p+1+\tau \right) }{1+\tau }\left\{
X_{0}B+a_{0}^{2}\right\} , \\
& \\
\widetilde{\widetilde{\Sigma }}& =\Delta \left[ a_{0}+AX_{0}\right]
^{2}+\delta \left\{ a_{0}^{2}+X_{0}B\right\} ^{2}.
\end{align*}

The determinant of the metric in the extended domain is expressed in the
form,

\begin{equation}
\mid g\mid =\frac{X}{4H}\left\{ 2\overline{A}\left( 2\overline{A}%
D+C^{2}\right) \left( \xi \zeta \right) ^{2}+\overline{B}\left( C^{2}\xi
\zeta -\overline{B}D\right) \right\} .
\end{equation}

In the limit as $\xi =0$ and/or $\zeta =0$ (which is equivalent to $\tau =1$
), the determinant becomes:

\begin{equation*}
\mid g\mid =\left[ \left( p+1\right) ^{2}+\left( l_{0}+a_{0}\sigma \right)
^{2}\right] ^{4}X_{0}^{2}\delta ^{2}.
\end{equation*}%
Note that the above determinant vanishes when $\sigma =\pm 1$. The points $%
\sigma =\pm 1$, however correspond to the null boundaries separating the
interaction region from the incoming regions ( see Fig. 2). Furthermore it
is clear from the equation $H=\delta -\Delta =\sin 2u\sin 2v$ that for $u=0,$
$0\leq v\leq \pi /2$ or $v=0,$ $0\leq u\leq \pi /2$ in either case $H=0$.
Therefore the extended metric (37) in ( $\xi ,\zeta $) coordinates includes
the space-time described in metric (30) exclusive of the null boundaries at $%
u=0,$ $v=0$. In other words, the metric (37) \ represents only the extension
of the interaction region.

The contravariant components of the metric (37) are

\begin{equation}
g^{\xi \xi }=\frac{\xi ^{2}\left( 4\widetilde{A}D+C^{2}\right) }{\left( 2%
\widetilde{A}\xi \zeta +\widetilde{B}\right) \left[ \left( 2\widetilde{A}%
D+C^{2}\right) \xi \zeta -\widetilde{B}D\right] },
\end{equation}

\begin{equation}
g^{\zeta \zeta }=\frac{\zeta ^{2}\left( 4\widetilde{A}D+C^{2}\right) }{%
\left( 2\widetilde{A}\xi \zeta +\widetilde{B}\right) \left[ \left( 2%
\widetilde{A}D+C^{2}\right) \xi \zeta -\widetilde{B}D\right] },
\end{equation}%
with%
\begin{align*}
\widetilde{A}& =\frac{\overline{A}}{2HX\delta }, \\
& \\
\widetilde{B}& =\frac{\overline{B}}{2HX\delta }.
\end{align*}

The nature of the surface when $\xi =\zeta =0$ is identified by calculating
the squared norms of the vector fields orthogonal to the surfaces $\xi _{0}=$
\textit{constant} and $\zeta _{0}=$ \textit{constant}$.$ Let the surface $%
S\left( \xi \right) =\xi -\xi _{0}$ be such that $\xi =\xi _{0}$ is not a
singular surface of the metric. The normal vector $N^{\mu }$ to the surface
is defined by $N^{\mu }=g^{\mu \nu }S,_{\nu }.$ Similarly for a surface $%
\zeta =\zeta _{0},$\ the norm squares are then obtained by%
\begin{eqnarray}
N^{2} &=&\left( \nabla S\right) ^{2}=g^{\mu \nu }\partial _{\mu }S\partial
_{\nu }S=g^{\xi \xi }, \\
&&  \notag \\
N^{2} &=&\left( \nabla S\right) ^{2}=g^{\mu \nu }\partial _{\mu }S\partial
_{\nu }S=g^{\zeta \zeta }.  \notag
\end{eqnarray}

It is clear to observe that the vector fields $\partial _{\mu }\xi $ and $%
\partial _{\mu }\zeta $ become null on the hypersurface $\xi =0,$ $\ \zeta
=0 $ respectively. Therefore the surface

\begin{equation}
\xi \zeta =\Delta \delta =0
\end{equation}%
consists of two null surfaces as depicted in Fig. 3. There are four distinct
regions assigned to the coordinates $\left( \xi ,\zeta ,r,\overline{y}%
\right) $. These \ regions are ; (i) $\xi >0,$ \ $\zeta >0$ which is part of
the interaction region $I_{0}$; \ (ii) $\xi <0,$ \ $\zeta <0$ isometric
region to $I_{0}$, since the simultaneous change in the sign of $\xi $ and $%
\zeta $ leaves the metric (37) invariant; (iii) the region for $\xi <0$, $%
\zeta >0$ and (iv) for $\xi >0$, $\zeta <0$.

From the curvature scalars as given in equations (18 and 19), the only
possibility that the curvature singularity may develop when,

\begin{equation}
l+a\sigma =0\text{ \ \ \ \ \ \ \ \ and \ \ \ \ \ }1+k\tau =0\text{\ \ \ \ }
\end{equation}%
occur simultaneously. This leads to $\tau =-\frac{1}{k}$ which is physically
not acceptable since the range of $\tau $ is $0\leq \tau \leq 1$, which is
positive definite. Even if $k<0$ is chosen ( i.e. negative root from
Eq.(17)) we can use the freedom of the NUT parameter $l,$ so that, for $l>a$
implies $l+a\sigma \neq 0,$ because the range of $\sigma $ is $-1\leq \sigma
\leq 1$ $\ $which makes $\mid g\mid \neq 0$. This is in marked distinction
from the case of CX for which $\mid g\mid =0$, in the extended domain. Thus,
the determinant of the metric together with the curvature scalars remain
bounded in the regions when $\xi <0$, $\zeta >0$ or $\xi >0$, $\zeta <0$. As
a result, the extended domain is free of any kind of singularity.

The geodesic description of CGW spacetime is almost well-known by now \cite%
{MT,HK1,DAL,MH} Any geodesic originating from the flat region fall into the
singularity ( or horizon) in a finite proper time. Some geodesics remain in
the incoming regions without ever crossing into the interaction region $%
I_{0} $. Depending on the initial conditions a vast majority of time-like
geodesics cross into the region $I_{0}$ and hit the horizon $\tau =1,$ in
our present case. In this sense we can refer to our CGW spacetime as
geodesically complete. We must admit, however, that the status of some
singularities ( i.e. fold or quasi-regular type) is still not well -
understood. There are even attempts to resolve this type of singularities by
using quantum mechanics (see for example Ref. \cite{HK2} and references
therein). This issue, as well as the stability of horizons is beyond our
scope and should better be addressed in a mathematical context. The
hypersurface $\xi =0$ $(\zeta =0)$ is a null surface and behaves like the
event horizon of a black hole ( one way membrane). This implies that the
future directed time-like or null trajectories originating from region $%
I_{0} $ can enter the region $\xi >0$, $\zeta <0$ or $\xi <0$, $\zeta >0$
and from these regions to the isometric region $I_{e}$ and continue in
replica regions.

\section{Colliding (\textsc{p}+2) - Forms in (\textsc{p}+4) Dimension.}

The action for the (4+p) - dimensional branes is given by

\begin{equation}
S=\int d^{4+p}x\sqrt{-g}\left( R-F_{\left( 2\right) }^{2}\right) ,
\end{equation}%
in which $F_{\left( 2\right) }$ \ represents the em 2-form. With reference
to the black p-brane solution \cite{GHT}.

\begin{equation}
ds_{4+p}^{2}=A\cdot B^{\frac{1-p}{1+p}}dt^{2}-(A\cdot B)^{-1}dr^{2}-B^{\frac{%
2}{p+1}}\sum_{i=1}^{p}\left( dy^{i}\right) ^{2}-r^{2}d\Omega _{2}^{2},
\end{equation}%
where

\begin{eqnarray*}
A &=&1-\frac{r_{+}}{r}, \\
B &=&1-\frac{r_{-}}{r},
\end{eqnarray*}%
in which $r_{+}$ and $r_{-}$ with $r_{+}>r_{-}>0,$ are the outer and inner
horizons, respectively. The corresponding CGW solution can easily be found 
\cite{HHU}. This is exactly the CGW solution in the higher dimensional EM
theory where the non zero component of the em 2-form is given by

\begin{equation}
F_{2}=Q\sin \theta d\theta \wedge d\varphi ,
\end{equation}%
in which $Q$ stands for the charge. Aplying duality on the 2-form in (p+4) -
dimension gives (p+2) - forms and one can obtain the collision problem of
the (p+2) - forms in (p+4) - dimensional space-time. Our (p+2) - form field
is defined in accordance with

\begin{equation}
\widetilde{F}^{\mu _{1}\mu _{2}...\mu _{p+2}}=\frac{1}{2}\mid g_{4+p}\mid
^{-1/2}\epsilon ^{\mu _{1}...\mu _{p+4}}F_{\mu _{p+3}\mu _{p+4}}.
\end{equation}

The new action takes the form

\begin{equation}
S=\int d^{4+p}x\sqrt{-g}\left( R-\frac{2}{\left( p+2\right) !}\widetilde{F}%
_{p+2}^{2}\right)
\end{equation}%
so that the field equations are

\begin{equation}
R_{\mu \nu }=\frac{2}{\left( p+2\right) !}\left( \widetilde{F}_{\mu \mu
_{1}...\mu _{p+1}}\widetilde{F}_{\nu }^{\mu _{1}...\mu _{p+1}}-\frac{\left(
p+1\right) }{\left( p+2\right) ^{2}}g_{\mu \nu }\widetilde{F}_{\mu
_{1}...\mu _{p+2}}\widetilde{F}_{\nu }^{\mu _{1}...\mu _{p+2}}\right)
\end{equation}

\begin{equation*}
\partial _{\mu }\left( \mid g_{4+p}\mid ^{1/2}\widetilde{F}^{\mu \mu
_{1}...\mu _{p+1}}\right) =0.
\end{equation*}

The solution can be expressed by

\begin{equation}
ds_{4+p}^{2}=\left( k+\tau \right) ^{2}\left( 2dudv-\delta dz^{2}\right)
-\left( \frac{1+\tau }{k+\tau }\right) ^{\frac{2}{p+1}}\sum_{i=1}^{p}\left(
dy^{i}\right) ^{2}-\left( \frac{1-\tau }{k+\tau }\right) \left( \frac{1+\tau 
}{k+\tau }\right) ^{\frac{1-p}{1+p}}dx^{2}
\end{equation}%
where

\begin{equation*}
k=\frac{r_{+}+r_{-}}{r_{+}-r_{-}}>1.
\end{equation*}

The em 2-form field is

\begin{equation}
F=Q\sqrt{\delta }\left[ -a\theta \left( u\right) du+b\theta \left( v\right)
dv\right] \wedge dz
\end{equation}%
while its dual is obtained in accordance with (47). We observe now that

\begin{equation}
\det \mid g_{4+p}\mid =\Delta \delta \frac{\left( k+\tau \right) ^{5}}{%
1+\tau }
\end{equation}

which indicates a coordinate singularity at $\tau =1.$ We introduce new
coordinates $\widetilde{l},r,$ and $x$ as we have done in the previous
chapters,

\begin{eqnarray}
\widetilde{l} &=&\sqrt{\Delta \delta } \\
r &=&\sin ^{2}u-\sin ^{2}v  \notag \\
\xi &=&\widetilde{l}e^{cx}  \notag \\
\zeta &=&\widetilde{l}e^{-cx}  \notag
\end{eqnarray}%
where c$^{2}=2^{\frac{1-p}{1+p}}$ $\left( k+1\right) ^{-2\left( \frac{p+2}{%
p+1}\right) }.$

Our new metric takes the form,

\begin{eqnarray}
ds_{4+p}^{2} &=&\frac{\left( k+\tau \right) ^{2}}{8H\widetilde{l}^{2}}%
\left\{ \left( \zeta d\xi +\xi d\zeta \right) ^{2}-\frac{2H\left( 1-\tau
\right) }{c^{2}\left( k+\tau \right) ^{3}\widetilde{l}^{2}}\left( \frac{%
1+\tau }{k+\tau }\right) ^{\frac{1-p}{1+p}}\left( \zeta d\xi -\xi d\zeta
\right) ^{2}\right\} \\
&&-\frac{\left( k+\tau \right) ^{2}}{2H}dr^{2}-\delta \left( k+\tau \right)
^{2}dz^{2}-\left( \frac{1+\tau }{k+\tau }\right) ^{\frac{2}{1+p}%
}\sum_{i=1}^{p}\left( dy^{i}\right) ^{2}  \notag
\end{eqnarray}

The determinant of the new metric expressed in the coordinates $\left( \xi
,\zeta ,r,z,y^{i}\right) $ is found to be

\begin{equation}
\mid g\mid ^{1/2}=\frac{\left( k+\tau \right) ^{2}}{4cH}
\end{equation}%
which is free of the coordinate singularity. It is observed that the
analytic extension does not effect the higher dimensional coordinates $%
\left( y^{i}\right) .$

\section{CONCLUSION.}

The physical significance of the NUT parameter in the theory of relativity
has been one of the discussion subject that its exact physical
interpretation is not clarified yet. In this paper, we have analysed the
effect of the NUT parameter in the context of CGW spacetimes. In our
analysis we considered the NUT extension of Paper II which is locally
isometric to the part of a region in between the inner and outer horizons of
KNN BH. Our main concern in this article is not only to provide a new CGW
geometry but also to explore the physical effect of the NUT parameter. We
have shown that the NUT parameter provides an additional twist to
gravitational waves. As a result, the waves that participate in the
collision is modified with respect to the cases in Paper I and II. The
overall effect of this modification becomes more clear when a non-unique
extension beyond the Cauchy horizon is obtained. The initial data of CX (in
Paper I and II) , however, was able to transform the coordinate singularity
into a harmless time-like singularity in the extended domain. In our case,
we have shown that the modification in the \ initial data as a result of the
inclusion of the NUT parameter removes the time-like singularities as well
and leaves the extended spacetime singularity free. We prove also that our
metric and the one considered previously by PX are distinct transforms
obtained from the most general PD family of solutions. Similar results to
ours were obtained by PX without identifying the role of the NUT parameter.
Similar technique applies also to the Cauchy-horizon forming higher
dimensional spaces.

To conclude the paper we wish to express the view that the CX duality is
more than a mathematical equivalence: It must address deeper implications
concerning BHs and colliding waves that awaits yet to be explored.

\begin{center}
Figure Captions
\end{center}

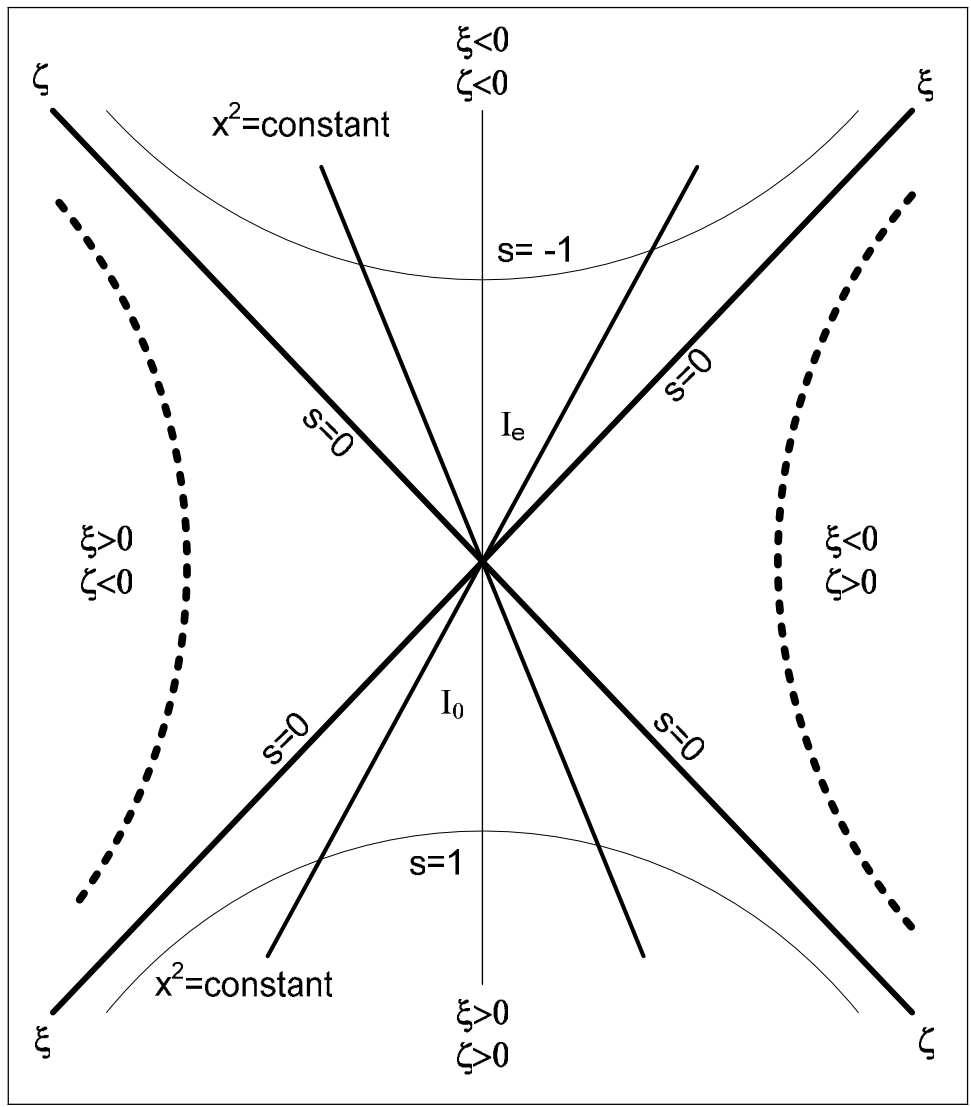
Figure 1: The manifold that represents analytically extended domain in the ($%
\xi ,\zeta ,r,x^{1}$) coordinates can be projected in the ( $\xi ,\zeta $ )
coordinates as above. There are four distinct regions. These regions are: i) 
$\xi >0,$ $\zeta >0$ which is part of the interaction region $I_{0}$; ii) \ $%
\xi <0,$ $\zeta <0$ isometric region $I_{e}$ to $I_{0}$, since the
simultaneous change in the sign of $\xi $ and $\zeta $ leaves the extended
metric invariant; iii) the region when $\xi >0,$ $\zeta <0$, and iv) and the
region $\xi <0,$ $\zeta >0$ are the regions that contain time-like
singularities along hyperbolic arcs shown with dashed lines.

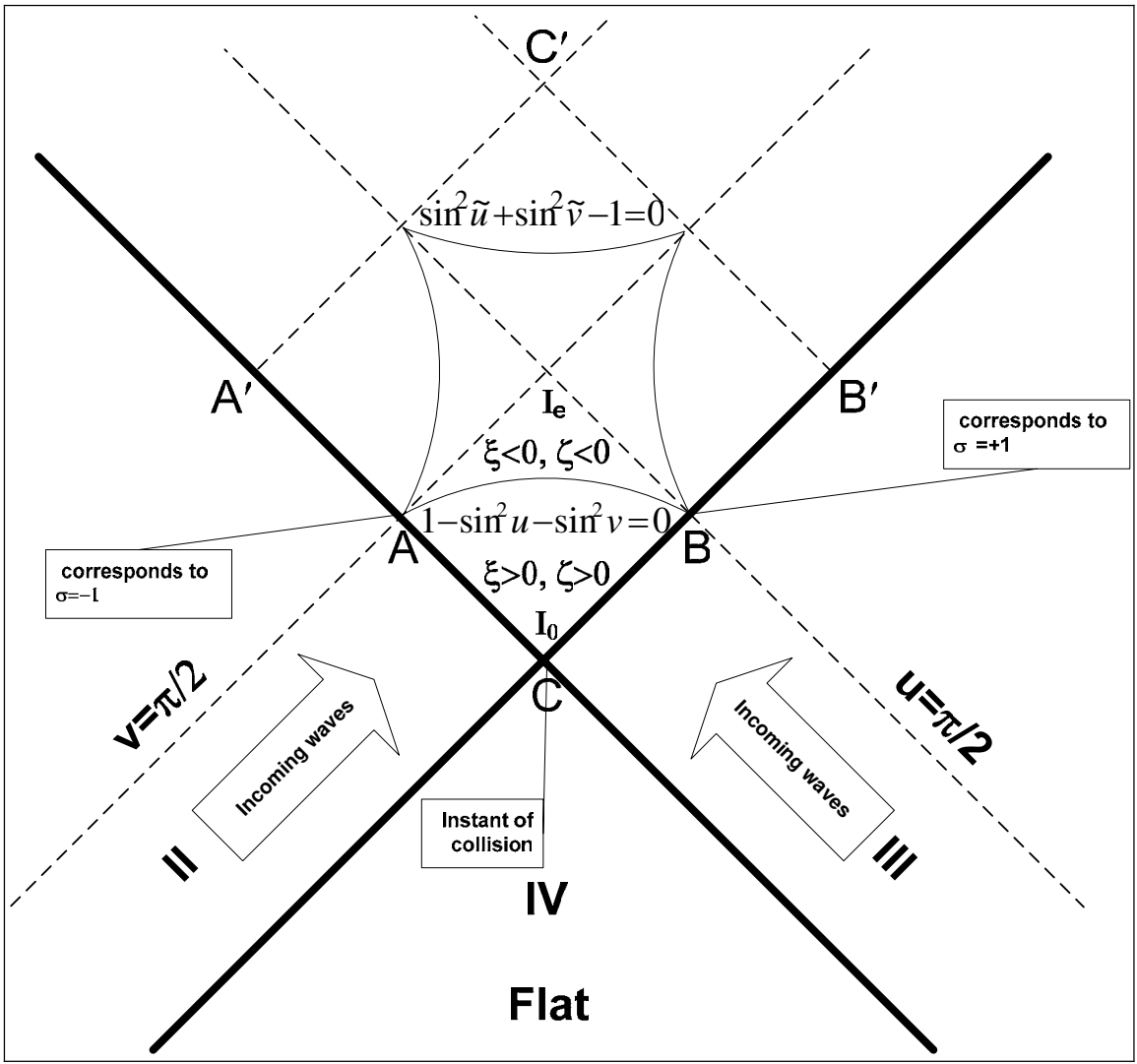
Figure 2: The space-time diagram describes the collision of Einstein -
Maxwell fields. Region II and III are the regions that contain the incoming
waves which are composed of plane impulsive gravitational wave accompanied
with shock gravitational and electromagnetic waves. Region IV is the flat
region before the arrival of the waves. The collision occur at point C.
Region I$_{\text{0}}$ is part of the interaction region. The arc AB is a
null hypersurface which is called Killing-Cauchy horizon that occurs at $%
\tau =1$ in which the analytic extension is performed. I$_{e}$ is the mirror
image of the region I$_{0}$. In the problem considered, the structure of the
extended domain is similar except the absence of time-like singularities
along hyperbolic arcs. The maximal analytic extension can be obtained
similar to paper I by tilling the entire region by panels which are replicas
of the panel CA'C'B'. \ 

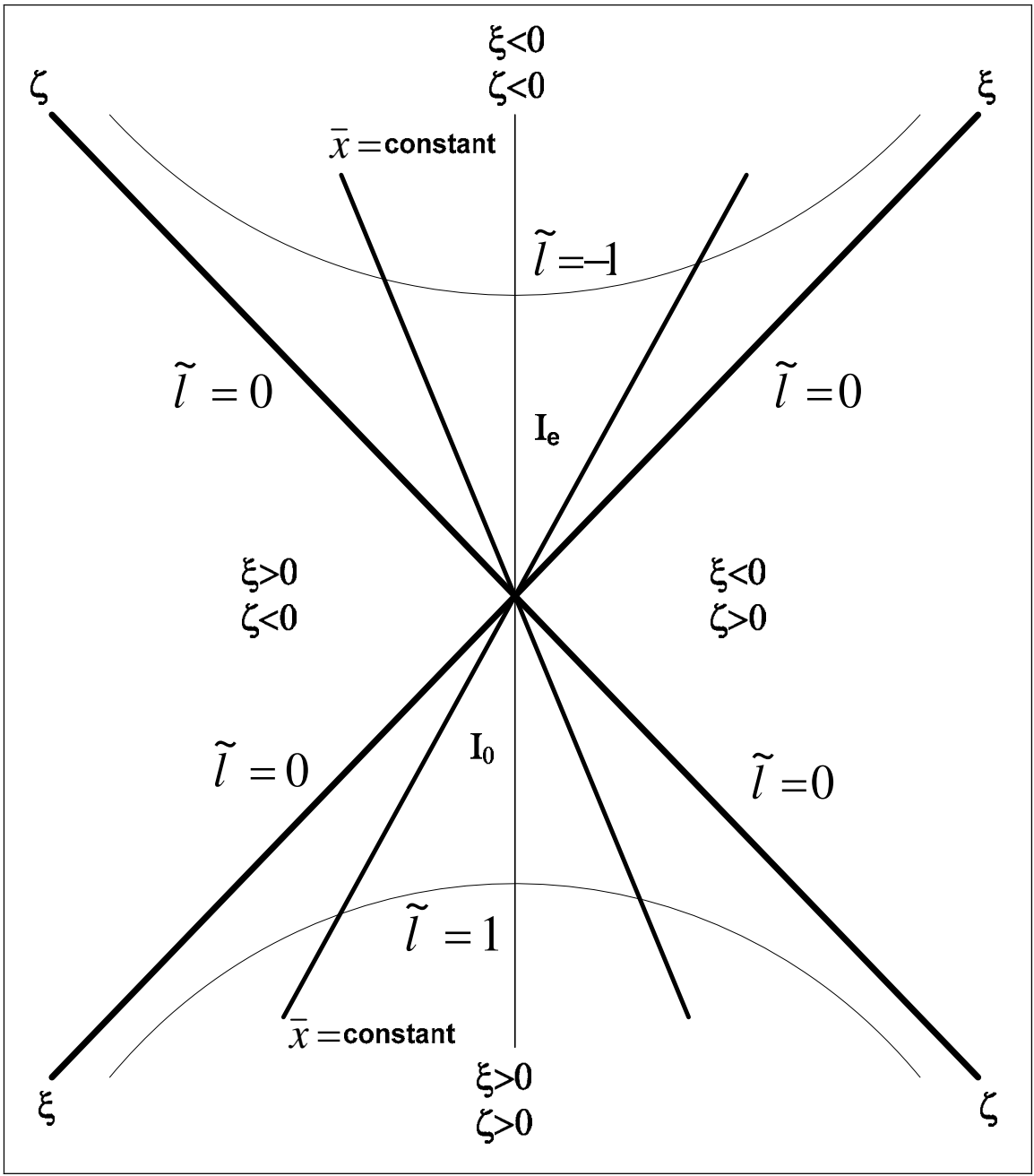
Figure 3: The space-time manifold in the ( $\xi ,\zeta ,r,\overline{y}$)
coordinates can be projected in the ( $\xi ,\zeta $) coordinates as above.
The hypersurface $\xi =\zeta =0$ are the null surfaces. By choosing a
variety of initial data and manipulating the symmetry ( i.e. taking the
combination of Killing vectors), the time-like singularities along
hyperbolic arcs can be eliminated and leaves the regions $\xi >0,$ $\zeta <0$
and $\xi <0,$ $\zeta >0$ singularity free. The regions I$_{0}$ and I$_{e}$
correspond to part of the interaction and the extended mirror image regions
when $\xi >0,$ $\zeta >0$ and $\xi <0,$ $\zeta <0$ respectively. The curves
of constant $\widetilde{l\text{ }}$are hyperbolae as in the case of paper I.

\end{document}